\documentstyle{l-aa}
\input psfig.tex
\ifx\undefined\AALPLAIN \else
\def\.#1{{\accent"C7 #1}}
\def\dot{\mathaccent"70C7 }
\fi

\begin{document}

\thesaurus{02.01.2, 11.01.2, 11.17.3, 13.25.3}

\title{UV to X-ray Spectra of Radio-Quiet Quasars.
Comparison With Accretion Disk Models}
\author{H. Brunner\inst{1}
\and    C. M\"uller\inst{2}
\and    P. Friedrich\inst{1}
\and    T. D\"orrer\inst{2}
\and    R. Staubert\inst{2}
\and    H. Riffert\inst{3}
}

\offprints{H. Brunner}

\institute{Astrophysikalisches Institut Potsdam,
An der Sternwarte 16, D-14482 Potsdam, Germany
\and
Institut f\"ur Astronomie und Astrophysik, Astronomie,
Universit\"at T\"ubingen, Waldh\"auserstr. 64, D-72076 T\"ubingen, Germany
\and
Institut f\"ur Astronomie und Astrophysik, Theoretische Astrophysik,
Universit\"at T\"ubingen, Auf der Morgenstelle, D-72076 T\"ubingen, Germany
}

\date{Received May 29; accepted June 12, 1997}

\maketitle
\begin{abstract}
We present UV to soft X-ray spectra of 31 radio-quiet quasars,
comprising data from the IUE ULDA database and the ROSAT
pointed observation phase. 90~\% of the sample members show a soft X-ray 
excess above an underlying hard X-ray power law spectrum. Particularly for 
the steep X-ray spectrum ($\alpha_{\rm energy} > 1.5$), low redshift subsample 
(17 objects) the X-ray spectral power law index is strongly correlated 
with the optical to X-ray broad-band spectral index $\alpha_{\rm ox}$, 
indicating 
that the main contribution to the soft X-ray and optical emission is due to 
the same emission component. We model the UV/soft X-ray spectra in terms of 
thermal emission from a geometrically thin $\alpha$-accretion disk. The 
structure and radiation field of the disk is calculated self-consistently and 
Compton scattering is treated by the Kompaneets equation. All relativistic 
effects on the disk structure and the emergent disk spectrum are included. 
Satisfactory spectral fits of the UV and soft X-ray continuum spectra are 
achieved when additional non-thermal hard X-ray and IR power law emission 
components are taken into account. The UV and soft X-ray spectra are well 
described by emission resulting from accretion rates in the range $\sim$ 
0.1 to $\sim$ 0.3 times the Eddington accretion rate. Low mass/low redshift
objects are found to accret at $< 0.15 \; \dot{M}_{\rm Eddington}$. 
Correlations of the accretion disk parameters with $\alpha_{\rm ox}$ are 
discussed.

\keywords{accretion disks -- active galaxies -- quasars -- X-rays} 
\end{abstract}

\section{Introduction}

It is a well established fact that the maximum of the
spectral energy distribution of quasars occures in the largely 
unobservable spectral range between the extrem UV and the soft 
X-ray domain. In many objects a steepening of the spectrum in the
soft X-ray range as compared to the hard X-rays is observed which, 
in combination with the turnover of
the spectrum in the UV range, suggests that these two components combine
in the largely unobservable range between $\sim 3 \times 10^{15}$ Hz and  
$\sim 3 \times 10^{16}$ Hz to form the so called big blue bump 
emission. Continued attempts have been made to derive a self-consistent
emission model which is able to account for this spectral component,
the most probable scenario being thermal emission from an accretion
disk around a central super-massive compact object.

The theory of standard geometrically thin $\alpha$-accretion disks
is largely based on the paper of Shakura \& Sunyaev (\cite{shakura}) and
a general relativistic version presented by Novikov \& Thorne (\cite{novikov}).
It has soon turned out that simple accretion disk models based on 
multi-temperature blackbody emission from an optically thick accretion disk
(Malkan \& Sargent \cite{malkan}, Malkan \cite{malkan2}), 
at sub-Eddington accretion rates, 
are not sufficiently hot, or else that highly super-Eddington accretion 
rates would be required, for the accretion disk to emit an appreciable 
fraction of the radiation in the soft X-ray range (Bechtold \cite{bechtold}). 
A number of authors have improved this simple model by considering 
various effects on the structure and emission spectrum of the disk.
Czerny \& Elvis (\cite{czerny}), and Wandel \& Petrosian (\cite{wandel})
calculated the radiative transfer by
including free-free opacities and the effects of Comptonization 
in a simple analytic manner.
Bound-free opacities as well as relativistic effects were 
included by Laor \& Netzer (\cite{laor89}) and Laor et al. (\cite{laor90}).
Most computations of model spectra to date, however, adopted 
a given vertical structure or made use of an averaging in the vertical
direction.
A detailed investigation of the emission spectrum was performed
by Ross, Fabian, \& Mineshige (\cite{ross}), using the Kompaneets equation 
(Kompaneets \cite{kompaneets}) to treat 
Compton scattering and including free-free and bound-free 
opacities of hydrogen and helium.
They solved the vertical temperature structure
and atomic level populations for a predetermined
constant vertical density profile. 
A self-consistent solution of the vertical structure 
and radiative transfer is given by 
Shimura \& Takahara (\cite{shimura}) and Shimura \& Takahara (\cite{shimura2})
for a Newtonian disk. In their viscosity description,
they made the ad hoc assumption that the local heating rate is
proportional to the mass density. 
In our approach (D\"orrer et al.\ \cite{doerrer}), the vertical structure and 
radiation field of a disk around a Kerr black hole is calculated 
in a self-consistent way. Moreover, we use a different viscosity description.
In section 4 a short review of this model is given. 

In the framework of the unified model the different
properties of Active Galactic Nuclei are explained as being due to the
different inclination angles under which the observer sees the
accretion disk as well as various additional components such as absorbing
material, emisssion line clouds or jets. The emission from the accretion 
disk is best studied in objects seen under intermediate inclination angles 
where the disk emission is neither obscured by an absorbing gas and dust 
torus nor is it swamped by beamed emission from a relativistic jet (in 
systems seen nearly face on). Such intermediate inclination angles are 
thought to lead to source properties as observed in radio-quiet quasars and 
Seyfert I galaxies.
Thus, high signal-to-noise data, covering the spectral range from the 
UV to the soft X-ray range, of samples of such objects are best suited to 
investigate whether their broad-band spectra may be understood in terms of 
emission from an accretion disk.   
We have selected a sample of 31 radio-quiet quasars (see section 2 for
a discussion of the selection criteria), which were observed both by IUE in 
the energy range from 130 to 305~nm and by ROSAT in the energy range from 
0.1~keV ($\sim$ 3~nm) to 2.4~keV ($\sim$ 0.5~nm) in order to investigate 
whether the UV and soft X-ray spectra of these objects are in agreement with 
predictions based on our accretion disk emission model. 

The X-ray emission in the ROSAT energy band consists of at 
least two components, a hard power law component, extending to higher X-ray
energies, beyond the ROSAT energy range, and a soft component, known under 
the name of soft X-ray 
excess emission, which in many AGN dominates the spectrum at energies below 
$\sim$ 0.5~keV and which is widely thought to originate in the inner
part of an accretion disk. 
Testing the predictions of our accretion disk model thus requires to 
separate the contributions of these two emission components. 
Due to the limited energy resolution of ROSAT and also due to its limited
spectral coverage at higher X-ray energies ($<$ 2.4~keV), except for the 
brightest objects, this is not possible based on the ROSAT data alone.
We therefore make use of published hard X-ray spectral slopes to separate the 
two emission components in our spectral fits.
As a first approach we have compared spectra from deep ROSAT
pointings with the hard power law spectra taken from the literature, to 
convince ourselves that soft excess emission is indeed present in almost all 
of the objects (see section 3).

Things are further complicated by the fact that in the UV range different
spectral components contribute to the emission. We treat this by including an 
additional power law component in our model fits extending from the IR with 
an exponential cutoff at around $2 \times 10^{15}$ Hz. 
Details on the model fitting performed and on the resulting distributions of
the four accretion disk model parameters central mass $M$, 
mass accretion rate $\dot{M}$, viscosity parameter $\alpha$, and the 
inclination angle $\Theta$ can be found in section 5.
A similar study using data from the ROSAT All Sky Survey based on
an earlier version of the present accretion disk model is
presented by Friedrich et al. (1997). A brief comparison of the
two models and a summary of basic results is given in Staubert et
al. (1997).

\begin{table*}
\caption [ ]{\label{sample} The AGN Sample\newline
ROR: ROSAT sequence number (PSPC). T is the exposure time in ks, z is
the redshift of the object, and R the radio flux in mJy. The UV flux of IUE 
(LWP and SWP camera) is given in Jy. CR is the ROSAT PSPC count rate [cts/s]
in the energy range 0.1 to 2.4~keV. See section 2 for details on the 
broad-band spectral indices $\alpha_{\rm ox}$.}
\begin{flushleft}
\begin{tabular}{rl|cc|rrccccccc}
\noalign{\smallskip}
\hline
\noalign{\smallskip}
 & AGN & RA & DEC &
\multicolumn{1}{c}{ROR} & 
\multicolumn{1}{c}{T} & z & R & m$_v$ &
\multicolumn{2}{c}{UV-flux} & CR & $\alpha_{\rm ox}$    \\
\# & name & (2000) & (2000) & & & & & & LW & SW & & \\
\noalign{\smallskip}
\hline \hline
\noalign{\smallskip}
1 & PG 0026$+$12      &  00 29 08 &  13 16 50 
& WG701456 &  2.73 & 0.142 & 2  & 15.4 & 1.02 & 1.08 & 0.413 & 1.381 \\
2 & PG 0052$+$251     &  00 54 47 &  25 26 19 
& WG701038 &  6.77 & 0.155 & 1  & 15.4 & 1.55 & 1.41 & 0.600 & 1.361 \\
3 & TON S210          &  01 21 43 & -28 20 01 
& US700445 &  4.53 & 0.117 & -  & 14.9 & -    & 1.72 & 2.000 & 1.410 \\
4 & MARK 1014         &  01 59 44 &  00 24 08 
& WG700225 &  6.25 & 0.163 & 8  & 15.7 & -    & 0.99 & 0.209 & 1.607 \\
5 & NAB 0205$+$02     &  02 07 43 &  02 43 21 
& US700432 & 14.07 & 0.155 & 2  & 15.4 & 1.60 & 0.91 & 0.667 & 1.523 \\
6 & PG 0804$+$761     &  08 10 59  &  76 02 14 
& WG700470 &  3.77 & 0.100 & 2  & 15.1 & 3.60 & 2.60 & 1.194 & 1.316 \\
7 & IRAS 09149$-$6206 &  09 16 00 & -62 20 11 
& WG701484 &  1.89 & 0.057 & 16 & 13.6 & 3.22 & 0.96 & 0.405 & 1.741 \\
8 & HE 1029$-$1401    &  10 31 48 & -14 17 51 
& WG700461 & 13.59 & 0.086 & -  & 13.9 & 6.18 & 3.80 & 1.830 & 1.432 \\
9 & PG 1049$-$005     &  10 51 45 & -00 52 20 
& US700381 &  4.74 & 0.357 & -  & 16.0 & 0.88 & 0.37 & 0.044 & 1.677 \\
10 & Q 1101$-$264      &  11 03 19 & -26 46 10 
& WG900525 &  5.12 & 2.148 & -  & 16.0 & 2.41 & -    & 0.036 & 1.589 \\
11 & PG 1116$+$215     &  11 19 02 &  21 18 13 
& WG700228 & 24.59 & 0.177 & 3  & 15.1 & 3.36 & 3.17 & 0.994 & 1.491 \\
12 & GQ COM            &  12 04 34 &  27 53 15 
& WG700232 & 14.42 & 0.165 & 1  & 15.6 & -    & 0.90 & 0.400 & 1.432 \\
13 & PG 1216$+$069     &  12 19 15 &  06 37 45 
& US700021 &  3.46 & 0.334 & 4  & 15.7 & -    & 0.86 & 0.395 & 1.427 \\
14 & PG 1247$+$268     &  12 49 59 &  26 30 13 
& WG701173 &  2.19 & 2.041 & 1  & 15.7 & 0.49 & -    & 0.104 & 1.554 \\
15 & B 201             &  12 59 43 &  34 22 35 
& WG600164 & 17.17 & 1.375 & 13 & 16.9 & 0.34 & -    & 0.049 & 1.487 \\
16 & PG 1307$+$085     &  13 09 41 &  08 19 15 
& WG700229 &  7.69 & 0.155 & -  & 15.3 & -    & 1.37 & 0.524 & 1.503 \\
17 & PG 1309$+$355     &  13 12 11 &  35 14 29 
& WG701079 &  6.24 & 0.184 & 43 & 15.5 & -    & 0.59 & 0.260 & 1.686 \\
18 & IRAS 13349$+$2438 &  13 37 10 &  24 22 19 
& UK700553 &  3.32 & 0.107 & 7  & 15.0 & 0.70 & -    & 0.954 & 1.584 \\
19 & PG 1352$+$183     &  13 54 27 &  18 04 44 
& US700804 &  5.63 & 0.152 & -  & 15.7 & 0.97 & 0.74 & 0.403 & 1.481 \\
20 & PG 1407$+$265     &  14 09 15 &  26 17 36 
& US700359 &  3.23 & 0.947 & 8  & 15.8 & 1.01 & 0.51 & 1.074 & 1.251 \\
21 & PG 1415$+$451     &  14 16 52 &  44 55 18 
& US700805 &  7.43 & 0.114 & -  & 15.7 & -    & 0.44 & 0.335 & 1.657 \\
22 & PG 1416$-$129     &  14 18 58 & -13 11 09 
& WG700527 &  9.11 & 0.129 & 4  & 15.4 & 0.61 & 0.34 & 0.378 & 1.389 \\
23 & PG 1444$+$407     &  14 46 37 &  40 34 23 
& US701371 &  5.62 & 0.267 & -  & 15.8 & -    & 0.74 & 0.334 & 1.640 \\
24 & PG 1522$+$101     &  15 24 16 &  09 58 09 
& US701001 & 10.02 & 1.325 & -  & 16.0 & 0.96 & 0.44 & 0.026 & 1.775 \\
25 & PG 1543$+$489     &  15 45 20 &  48 45 37 
& US700808 &  7.18 & 0.400 & 1  & 16.3 & 0.64 & -    & 0.115 & 1.685 \\
26 & MARK  876         &  16 13 44 &  65 42 47 
& WG700230 &  6.72 & 0.129 & 2  & 15.4 & 1.66 & 1.32 & 0.915 & 1.354 \\
27 & MARK  877         &  16 20 03 &  17 24 14 
& UK700274 &  9.57 & 0.114 & 1  & 15.4 & 1.65 & -    & 0.117 & 1.578 \\
28 & PG 1630$+$377     &  16 31 52 &  37 37 30 
& WG700255 &  8.93 & 1.478 & -  & 16.0 & 0.93 & 0.23 & 0.073 & 1.604 \\
29 & HS 1700$+$6416    &  17 00 47 &  64 11 57 
& WG701457 & 27.41 & 2.722 & -  & 16.1 & 0.16 & -    & 0.012 & 1.775 \\
30 & KUV 18217$+$6419  &  18 21 44 &  64 20 44 
& US700948 &  1.98 & 0.297 & 13 & 14.2 & 3.76 & 2.40 & 1.358 & 1.416 \\
31 & MR 2251$-$178     &  22 53 59 & -17 33 50 
& UK701630 & 18.32 & 0.068 & 3  & 14.4 & 3.17 & 1.94 & 3.046 & 1.272 \\
\noalign{\smallskip}
\hline
\noalign{\smallskip}
\end{tabular}
\end{flushleft}
\end{table*}

\section{Sample selection and data analysis}

We have selected a sample of bright radio-quiet objects from the Veron-Cetty 
\& Veron (\cite{veron}) and the Hewitt \& Burbidge (\cite{hewitt}) quasar 
catalogues by requiring the radio (flux measured at 5~GHz) to optical 
(V band) broad-band spectral index $\alpha_{\rm ro}$ to be flatter than 0.3 . 
For objects where no radio flux measurement was available, an upper limit 
of 25~mJy was assumed. Note that, through this selection criterium, objects 
weaker than $16.8^{\rm th}$ magnitude in the v band are only included in the 
sample when radio flux measurements are available and the radio flux is 
below 25~mJy, essentially resulting in a cutoff of the sample at 
$m_{\rm v} = 16.8$ . All objects show broad emission lines and have absolute 
optical magnitudes brighter than M$_{\rm v} = -23.5$ and are thus 
classified as quasars.

These objects were cross-correlated both with the list of ROSAT
PSPC pointings from the ROSAT archive and with the IUE ULDA database of 
low resolution spectra. Objects were included in the sample
when (1) at the  time of the analysis (summer 1995), UV spectra were available
in the ULDA database and the ROSAT data had 
entered the public domain, (2) the objects were located within 44 minutes of 
arc from the optical axis of the ROSAT observation, (3) they were not hidden 
by the PSPC support structure, (4) at least 180 PSPC source counts
had been measured.
The final sample defined in this way which was investigated in our further 
analysis consists of 31 objects, listed in Tab. \ref{sample}.
 
Depending on the distance of the object from the optical axis of the ROSAT 
PSPC detector, X-ray counts were extracted within a circular area centered on
the source position with  radii ranging from 1 to 5 minutes of arc.
For each object, the background was accumulated from an annulus centered
on the source position from which areas contaminated by other sources had been
excluded. The data were binned into background subtracted count
rate spectra with, depending on the number of source counts,  6 to 23 
spectral bins in the energy range from 0.1 to 2.4~keV (channel
11 to 240) and a dead time correction and a correction for telescope 
vignetting was applied. 
The binning was performed such that a S/N of at least 5 was achieved in
each bin.

The broad-band spectral indices $\alpha_{\rm ro}$ and $\alpha_{\rm ox}$ used
for the sample selection and in our statistical analysis were calculated
from the source frame luminosity densities at 5~GHz, 2500~\AA, and 2~keV, 
respectively, which were determined following Zamorani et al. 
(\cite{zamorani}) and Marshall et al. (\cite{marshall}), and applying 
corrections for reddening following Seaton (\cite{seaton}). The resulting
$\alpha_{\rm ox}$ values are listed in Table \ref{sample}.

Data from the IUE ULDA database were used to determine continuum flux 
values in each of the frequency bands 130--140, 140--155, 166--170, 
and 170--185~nm (SWP camera) and 245--260, 260--275, 275--290, 
and 290--305~nm (LWP/LWR camera). The continuum flux level was determined 
by removing emission and absorption lines using a sliding window technique 
with window widths of 0.4, 0.8, 1.6, 3.2, 6.4, and 12.8~nm. De-reddening of 
the resulting continuum fluxes was performed by using
${\rm <N_H/E_{B-V}>=4.8 \times 10^{21} \; cm^{-2} \; mag^{-1}}$ 
(Bohlin et al.\ \cite{bohlin})
with galactic ${\rm N_H}$ values as derived from 21~cm radio measurements  
(Stark et al.\ \cite{stark} and Elvis et al.\ \cite{elvis}).

\begin{figure*}
\par\centerline{\psfig{figure=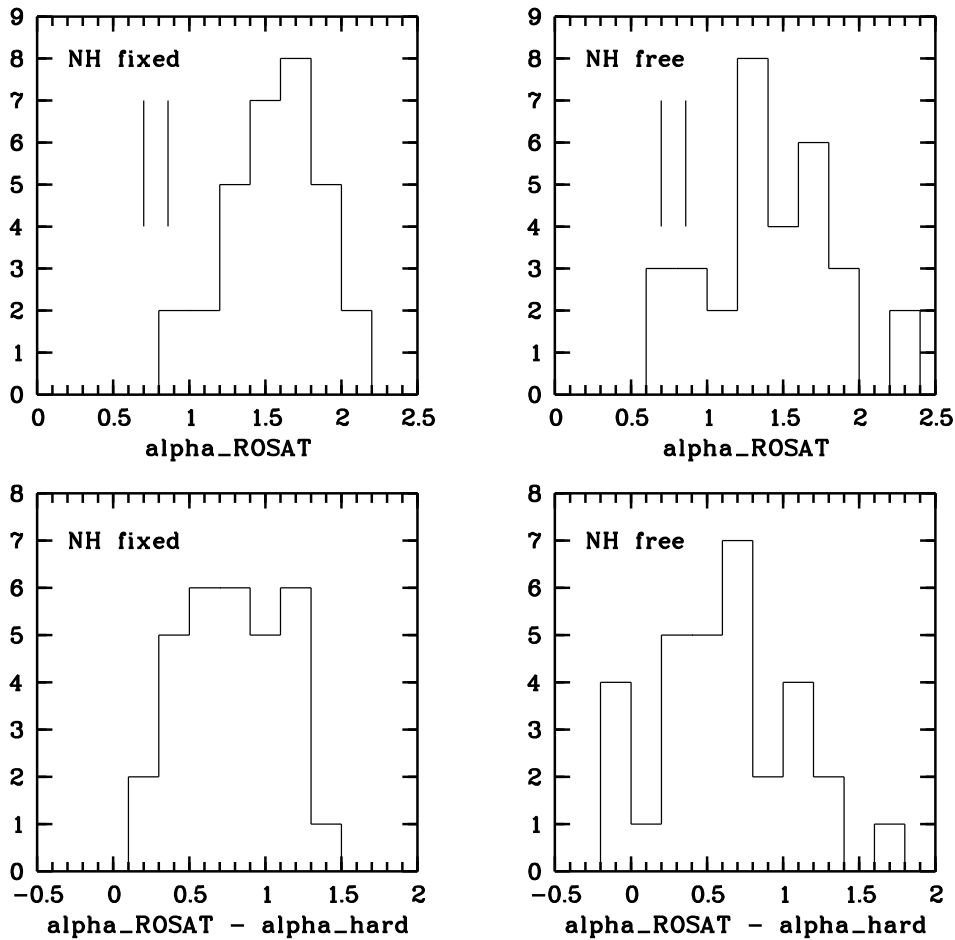,width=15cm,clip=}}
\caption[]{Histograms of ROSAT best-fit spectral power law indices.
The canonical hard X-ray spectral power law index, $\alpha=0.7$, as well as 
the mean hard X-ray spectral power law index, $\alpha=0.86$, of the sample 
members where measurements were available are marked by vertical lines.
Lower panels: Histograms of change in spectral slope between ROSAT and
hard X-ray band.}
\label{powerlaws}
\end{figure*}

\begin{figure*}
\par\centerline{\psfig{figure=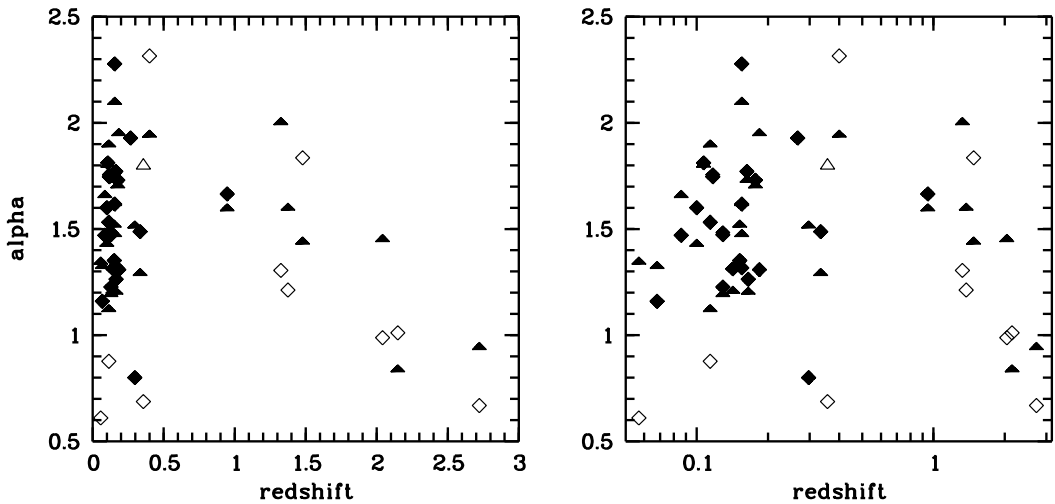,width=15cm,clip=}}
\caption[]{ROSAT spectral power law index plotted over redshift (left: linear
scale; right: logarithmic scale). Each object is shown twice. Diamonds: 
Fits with free ${\rm N_H}$. Triangles: Fits with fixed ${\rm N_H}$. Open 
symbols refer to spectral fits where the $1 \sigma$ statistical errors were 
larger than 0.3. 
\label{redsh}}
\end{figure*}

\begin{table}
\caption [ ]{${\rm N_H}$ in units of $10^{20} {\rm cm}^{-2}$.
The first ${\rm N_H}$ value (${\rm N_H}$ - fix) is taken from Elvis et al.
\cite{elvis} if marked with a $^{\star}$, else from Stark et al. 
\cite{stark}. $\alpha$ is the ROSAT spectral power law index, 
$\sigma _{\alpha}$ the corresponding $1 \sigma$ error.
$\sigma _{\rm N_H}$ ist the $1 \sigma$ error of the fittet ${\rm N_H}$ value.
\label{nh}}
\begin{flushleft}
\begin{tabular}{l|r@{.}lcc|cccc}
\hline \noalign{\smallskip}
\multicolumn{1}{c}{ } & \multicolumn{4}{|c|}{${\rm N_H}$ - fix} &
\multicolumn{4}{c}{${\rm N_H}$ - free}\\
\# & \multicolumn{2}{c}{${\rm N_H}$} & $\alpha$ & $ \sigma _{\alpha}$
& ${\rm N_H}$ & $\sigma _{\rm N_H}$ & $\alpha$ & $\sigma _{\alpha}$ \\
\hline \hline \noalign{\smallskip}
 1 &  4 & 93$^{\star}$ & 1.214 & 0.085 & 5.33 & 1.04 & 1.312 & 0.269\\
 2 &  4 & 50$^{\star}$ & 1.481 & 0.040 & 3.93 & 0.44 & 1.317 & 0.139\\
 3 &  1 & 67           & 1.771 & 0.026 & 1.61 & 0.19 & 1.746 & 0.093\\
 4 &  2 & 45           & 1.736 & 0.067 & 2.54 & 0.69 & 1.771 & 0.280\\
 5 &  2 & 99$^{\star}$ & 2.104 & 0.025 & 3.44 & 0.30 & 2.277 & 0.118\\
 6 &  3 & 12$^{\star}$ & 1.435 & 0.033 & 3.62 & 0.40 & 1.600 & 0.140\\
 7 & 17 & 96           & 1.351 & 0.270 & 7.15 & 2.45 & 0.610 & 0.386\\
 8 &  6 & 64           & 1.665 & 0.020 & 5.82 & 0.21 & 1.470 & 0.056\\
 9 &  3 & 97           & 1.798 & 0.330 & 0.98 & 1.07 & 0.687 & 0.769\\
10 &  5 & 25           & 0.844 & 0.269 & 6.24 & 6.37 & 1.011 & 0.739\\
11 &  1 & 44$^{\star}$ & 1.711 & 0.016 & 1.48 & 0.11 & 1.730 & 0.054\\
12 &  1 & 72$^{\star}$ & 1.210 & 0.030 & 1.86 & 0.29 & 1.263 & 0.117\\
13 &  1 & 47           & 1.298 & 0.059 & 1.92 & 0.55 & 1.488 & 0.238\\
14 &  1 & 57           & 1.458 & 0.173 & 0.52 & 0.81 & 0.988 & 0.571\\
15 &  1 & 90           & 1.605 & 0.115 & 1.01 & 0.85 & 1.213 & 0.422\\
16 &  2 & 20$^{\star}$ & 1.627 & 0.037 & 2.18 & 0.36 & 1.617 & 0.153\\
17 &  1 & 84           & 1.957 & 0.073 & 0.55 & 0.32 & 1.309 & 0.197\\
18 &  1 & 09           & 1.808 & 0.049 & 1.09 & 0.28 & 1.812 & 0.162\\
19 &  1 & 84$^{\star}$ & 1.526 & 0.051 & 1.43 & 0.39 & 1.352 & 0.182\\
20 &  1 & 38$^{\star}$ & 1.602 & 0.040 & 1.52 & 0.32 & 1.665 & 0.153\\
21 &  1 & 26           & 1.903 & 0.061 & 0.59 & 0.24 & 1.532 & 0.157\\
22 &  7 & 20$^{\star}$ & 1.198 & 0.066 & 7.37 & 0.80 & 1.227 & 0.155\\
23 &  1 & 11           & 1.943 & 0.066 & 1.08 & 0.37 & 1.928 & 0.207\\
24 &  3 & 02           & 2.008 & 0.264 & 1.21 & 1.10 & 1.305 & 0.799\\
25 &  1 & 59           & 1.949 & 0.101 & 2.38 & 0.88 & 2.315 & 0.416\\
26 &  2 & 66$^{\star}$ & 1.499 & 0.029 & 2.59 & 0.31 & 1.473 & 0.120\\
27 &  4 & 35$^{\star}$ & 1.128 & 0.110 & 3.43 & 1.08 & 0.877 & 0.328\\
28 &  0 & 90$^{\star}$ & 1.446 & 0.113 & 1.64 & 0.87 & 1.835 & 0.463\\
29 &  2 & 49           & 0.949 & 0.179 & 1.58 & 1.74 & 0.669 & 0.602\\
30 &  4 & 15           & 1.522 & 0.054 & 1.84 & 0.51 & 0.800 & 0.182\\
31 &  2 & 84$^{\star}$ & 1.331 & 0.010 & 2.36 & 0.10 & 1.160 & 0.037\\
\hline \noalign{\smallskip}
\end{tabular}
\end{flushleft}
\end{table}

\begin{figure}
\par\centerline{\psfig{figure=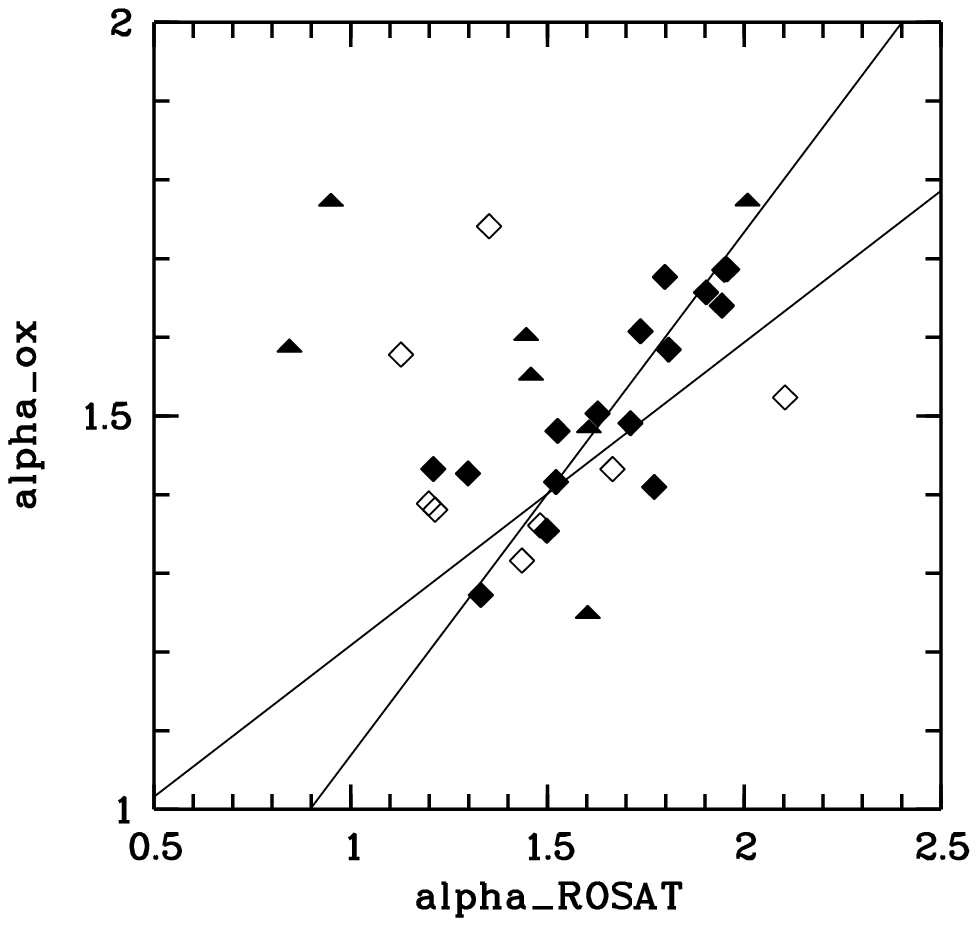,width=8.8cm,clip=}}
\caption[]{Broad band power law index $\alpha_{\rm ox}$ plotted against ROSAT
spectral power law index (fixed ${\rm N_H}$). Diamonds: redshift, $z<0.5$; 
triangles: $z>0.5$; open symbols: absorbed spectra 
(${\rm N_H}>3 \cdot 10^{20} \; {\rm cm}^{-2}$). 
$1 \sigma$ errors of the slope of the linear regression line calculated for
the 17 steep X-ray spectrum objects ($\alpha_{\rm ROSAT}>1.5$) are marked as 
solid lines. \label{aoxafix}}
\end{figure}

\section{Spectral power law fits to the ROSAT data}

As a first approach to quantify the soft X-ray excess emission in the
ROSAT energy window
we have fitted power law spectra to the ROSAT count rates for all sample
members. The distribution of the resulting spectral indices 
$\alpha_{\rm ROSAT}$ for a column density fixed at the galactic value and 
a free absorbing column density ${\rm N_H}$ are given in Fig. \ref{powerlaws} 
(upper panels). 
For comparison, the canonical value, $\alpha_{\rm hard} = 0.7$, and the sample 
mean, $<\alpha_{\rm hard}>$ = 0.86, of the hard X-ray power law are indicated 
as vertical lines.
Except for a somewhat larger width of the distribution no large
difference between the fixed, galactic ${\rm N_H}$ and free ${\rm N_H}$ 
power law spectral
indices is observed, thus indicating that, on average, the spectra are not
much affected by intrinsic low energy absorption in excess of the galactic
value in the ROSAT spectral range.
The spectral indices, together with the corresponding
column densities ${\rm N_H}$, are summarized in Table \ref{nh}. 
The mean ROSAT spectral power law indices for free and fixed ${\rm N_H}$ are
1.40 and 1.55, respectively, signifying a marked steepening of the
spectrum as compared to the spectral slopes observed at higher energies.
Looking at each object individually, a steepening of the
spectral slope between the hard and soft X-ray range is found in almost
all sample members. The mean change in spectral index is found to be 0.62 
(free ${\rm N_H}$) and 0.78 (fixed ${\rm N_H}$), respectively. In Fig. 
\ref{powerlaws} (lower panels) the distribution of this change in spectral 
index is shown. 

When going to objects at higher redshifts, the ROSAT sensitivity window
is shifted to higher source frame energies, thus in effect turning
ROSAT into a higher energy X-ray instrument. 
We find that at these higher source frame X-ray energies ROSAT does indeed 
measure harder X-ray spectral indices, similar to those of low redshift 
objects measured by higher energy X-ray instruments (see Fig. \ref{redsh}), in 
agreement with results by Schartel et al. (1996) and others. Note, however,
that Puchnarewicz et al. (\cite{puch96}) do not find a dependence of 
the mean X-ray power law index on redshift in their sample of AGN from 
the RIXOS survey which may in part be due to the selection of their objects
in the hard 0.4 -- 2.0 keV ROSAT energy band. 

Contrary to previous results, here, the dependence of the X-ray spectral
index on redshift is visible in individual, relatively bright objects as 
opposed to averaged properties derived from stacked spectra or mean spectral 
indices of many weak objects. Based on radio flux measurements and upper 
limits all objects in the sample are known to be radio-quiet ($\alpha_{ro}
< 0.3$; see section 2). We can therefore exclude any contamination of our 
sample from high-redshift, radio-loud objects which are known to be more
luminous than radio-quite quasars and at the same time display harder X-ray 
spectra. Such a contamination has previously been suggested as a possible
cause of the observed dependence of the ROSAT power law indices on
redshift. For the 7 objects in the redshift range $0.5 < z < 3.0$ a Spearman 
rank correlation coefficient of $-0.82 \; (-0.75)$ is found, corresponding to 
a likelihood of 0.02 (0.05) for randomness of the $\alpha_{\rm ROSAT} - z$ 
correlation. Results for fixed ${\rm N_H}$ and free ${\rm N_H}$ 
(in brackets) spectral power law fits are given.   
A quantitativ analysis of this behaviour, using the accretion disk model 
described in the following section is presented in a forthcoming paper
(Brunner et al., 1997). Note that the low redshift objects ($z < 0.5$) in our 
sample do not follow this trend, suggesting a turnover of the 
$\alpha_{\rm ROSAT} - z$ relation in the redshift range $0.2 < z < 0.5$ 
with decreasing spectral indices on either side of this range. 
One possible interpretation for the turnover of the $\alpha_{\rm ROSAT} - z$ 
relation is that, as one goes to lower and lower redshifts, 
lower luminosity objects are detected where the accretion disk component is 
increasingly absorbed and the spectrum is increasingly dominated by the 
hard power law component.
We would finally like to point out that the observed hardening of the ROSAT 
spectral indices with redshift rules out the possiblity that the steep AGN 
spectra observed by ROSAT may, as has been suggested, in part be due to 
errors in the cross-calibration of the ROSAT PSPC detector and previous 
higher energy X-ray instruments.
 
For the steep X-ray spectrum ($\alpha_{\rm ROSAT} \; {\rm (fixed N_H)} > 1.5$) 
subsample (17 objects) a strong correlation of the ROSAT spectral index 
(fixed ${\rm N_H}$) and the optical to X-ray broad-band spectral index is 
found (see Fig. \ref{aoxafix}). The correlation index is 0.78 (Spearman rank 
correlation coefficient), corresponding to a probability of $2.1 \cdot 10^{-4}$
of randomness. This suggests that in objects with strong soft X-ray excess 
emission, i.e., objects with steep ROSAT spectra, the dominant contributions 
to the X-ray and UV/optical emission are due to the same physical emission 
component (i.e., the big blue bump emission). While the correlation can also 
be traced to objects with lower ROSAT spectral indices, a number of these 
objects show broad-band spectral indices which are considerably steeper than 
predicted by the correlation, suggesting that in these objects, the onset of 
the big blue bump emission is at or below the lower cutoff of the ROSAT 
sensitivity window. Note that a fraction of these objects (marked by filled
triangles in Fig. \ref{aoxafix}) are at higher redshifts ($z>0.5$) where
any blue bump emission component is expected to be shifted out of the ROSAT 
sensitivity window. Objects marked as open diamonds in Fig. \ref{aoxafix}
which also do not seem to follow the correlation are seen through absorbing 
column densities $> 3 \cdot 10^{20}$ cm$^{-2}$ and may thus be affected by 
uncertainties in the ROSAT power law index (fixed ${\rm N_H}$) and possibly 
the de-reddening of the optical fluxes. Similar correlations of 
$\alpha_{\rm ox}$ and $\alpha_{\rm x}$ have also been reported by 
Puchnarewicz et al. 1996.

While the change in spectral slope between the ROSAT and harder
X-ray energy bands as well as the $\alpha_{\rm ROSAT} - \alpha_{\rm ox}$ 
correlation are useful indicators for the soft X-ray excess and big blue bump 
emission, a quantitative analysis is best performed 
in the framework of a physical emission model, the most 
widely advocated candidate  being emission from the hot inner region of 
an accretion disk around a super-massive central object.

\section{Accretion disk model used in spectral fits}

We assume a standard geometrically thin $\alpha$-accretion disk
around a massive Kerr black hole. 
A detailed description of our model is given in 
D\"orrer et al.\ (\cite{doerrer}).
Here, we just summarize the basic concepts of the calculations.
Model parameters are the mass $M$ of the central black hole, the 
accretion rate $\dot{M}$, the viscosity parameter $\alpha$, and  
the specific angular momentum $a$ of the central black hole.
In addition to the parameters describing the physical properties of the
accretion disks the inclination angle under which the observer sees the disk
is also a free parameter. The specific angular momentum
$a$ of the central black hole was fixed at $a=0$ in our spectral fits.
All relativistic corrections on the disk structure with respect to
a Newtonian model are included according to Riffert \& Herold 
(\cite{riffert}).

The radiative transfer is solved in the Eddington approximation,
and the plasma is assumed to be in a state of local thermodynamic
equilibrium.
Multiple Compton scattering is treated in the Fokker-Planck
approximation using the Kompaneets operator.
The absorption cross section contains only free-free processes
for a pure hydrogen atmosphere.
Induced contributions to the radiative processes have been neglected 
throughout.

For a given radial distance $R$ from the central black hole,
a self-consistent solution of the vertical structure and radiation field 
of the disk is obtained from the hydrostatic equilibrium equation,
radiative transfer equation, energy balance equation, and equation 
of state, when proper boundary conditions are imposed.
Note, that we have not considered convection in our model.

In our calculations the viscosity is assumed to be entirely 
due to turbulence. Because the standard $\alpha$-description 
(viscosity proportional to the total pressure) leads to 
diverging temperature profiles in the upper parts
of the disk, we have included the radiative energy loss of the turbulent
elements in the optically thin regime.
For the turbulent viscosity we then have

\begin{equation}
\eta = \alpha \; \rho \; H \; V_{\rm turb}~,
\end{equation}

where $\eta$ is the shear viscosity,
$\rho $ is the mass density, $H$ is the self-consistently calculated 
height of the disk, $V_{\rm turb}$ is the upper limit
for the velocity of the largest turbulent elements
(see D\"orrer et al.\ \cite{doerrer} 
for details on the determination of $V_{\rm turb}$)
and $\alpha$ is the viscosity parameter of our model.

We used a finite difference scheme in the vertical direction $z$
and in frequency space $\nu$ to find solutions of the given set of equations.
The vertical structure was resolved with 100 points on a
logarithmic grid, and 64 grid points were used in frequency space. 
The resulting set of algebraic difference equations was then solved by
a Newton-Raphson method.
 
To get the overall structure and emission spectrum of the disk
we calculated the vertical structure and the local emission spectrum
at $50$ logarithmically spaced radial grid points from the last stable orbit
to the outer disk radius (here $R_{\rm out}=1000 \; R_{\rm S}$, where 
$R_{\rm S}=2GM/c^{2}$ is the Schwarzschild radius, $G$ is the gravitational 
constant, and $c$ is the velocity of light).
The whole disk spectrum as seen by a distant observer at an inclination 
angle $\Theta _{0}$ with respect to the disk axis ($\Theta _{0}=0$
for a face-on observer) is then calculated by integration of the local
spectra over the disk surface. All general relativistic effects on the 
propagation of photons from the disk surface to the observer are 
included, using a program code (Speith et al. 1995) to obtain values
of the Cunningham transfer function (Cunningham \cite{cunningham})
for any set of parameters. 

The accretion disk spectrum as determined from the above calculations
extends from the optical to the soft X-ray range ($<$ 1 keV), the maximum 
of the emission being in the far UV. Qualitatively, the dependence of the 
spectral shape on the model parameters is such that, if the mass accretion 
rate $\dot{M}$ is kept at a constant fraction of the Eddington accretion 
rate $\dot{M}_{\rm Eddington}$, the central mass $M$ mainly determines the 
total flux while approximately maintainig the spectral shape. Increasing
the mass accretion rate in terms of the Eddington accretion rate 
$\dot{M}/\dot{M}_{\rm Eddington}$, on the other hand, while also increasing the
total flux, makes the spectrum harder, i.e. a larger fraction of the
flux is emitted in the X-ray range. Similarly, an increase of the viscosity
parameter $\alpha$ also leads to a hardening of the spectrum.  
Going from low (disk seen face on) to high (disk seen edge on) inclination 
angles of the accretion disk, the total flux from the disk is reduced and
at the same time a hardening of the spectrum due to Doppler boosting 
resulting from the rotation of the disk occures. 

\section{Model fits to the IUE and ROSAT data}

We have performed $\chi^2$ fits to the 
UV to soft X-ray continuum spectra of the 31 sample quasars in order 
to investigate whether the observed spectra are indeed in agreement
with our model, i.e. whether satisfactory best-fit minimum $\chi^2$ values
are achieved and the resulting accretion disk model parameters
are in line with our general understanding of the quasar phenomenon.

\begin{figure}
\par\centerline{\psfig{figure=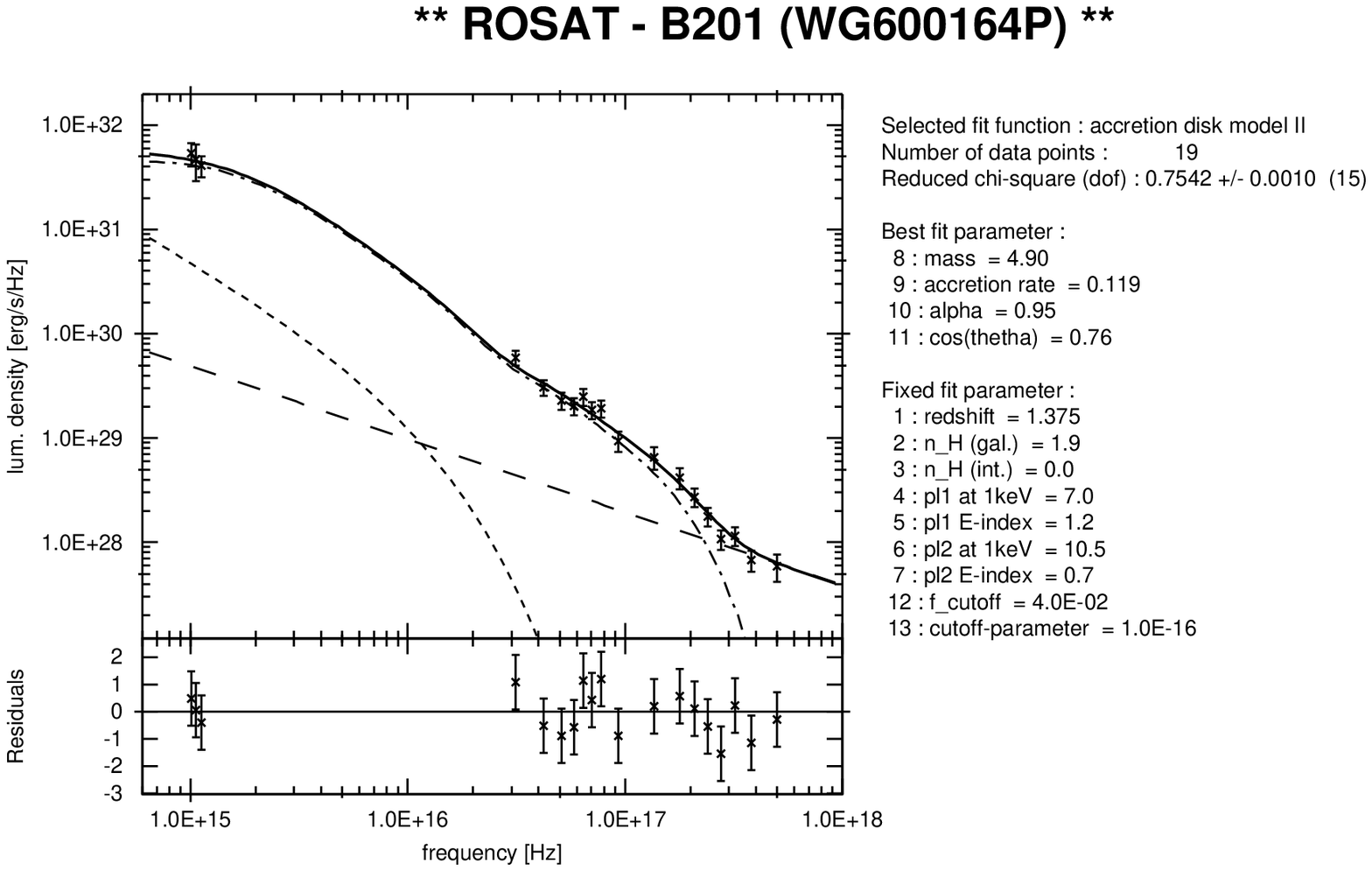,width=8.0cm,clip=}} 

\par\centerline{\psfig{figure=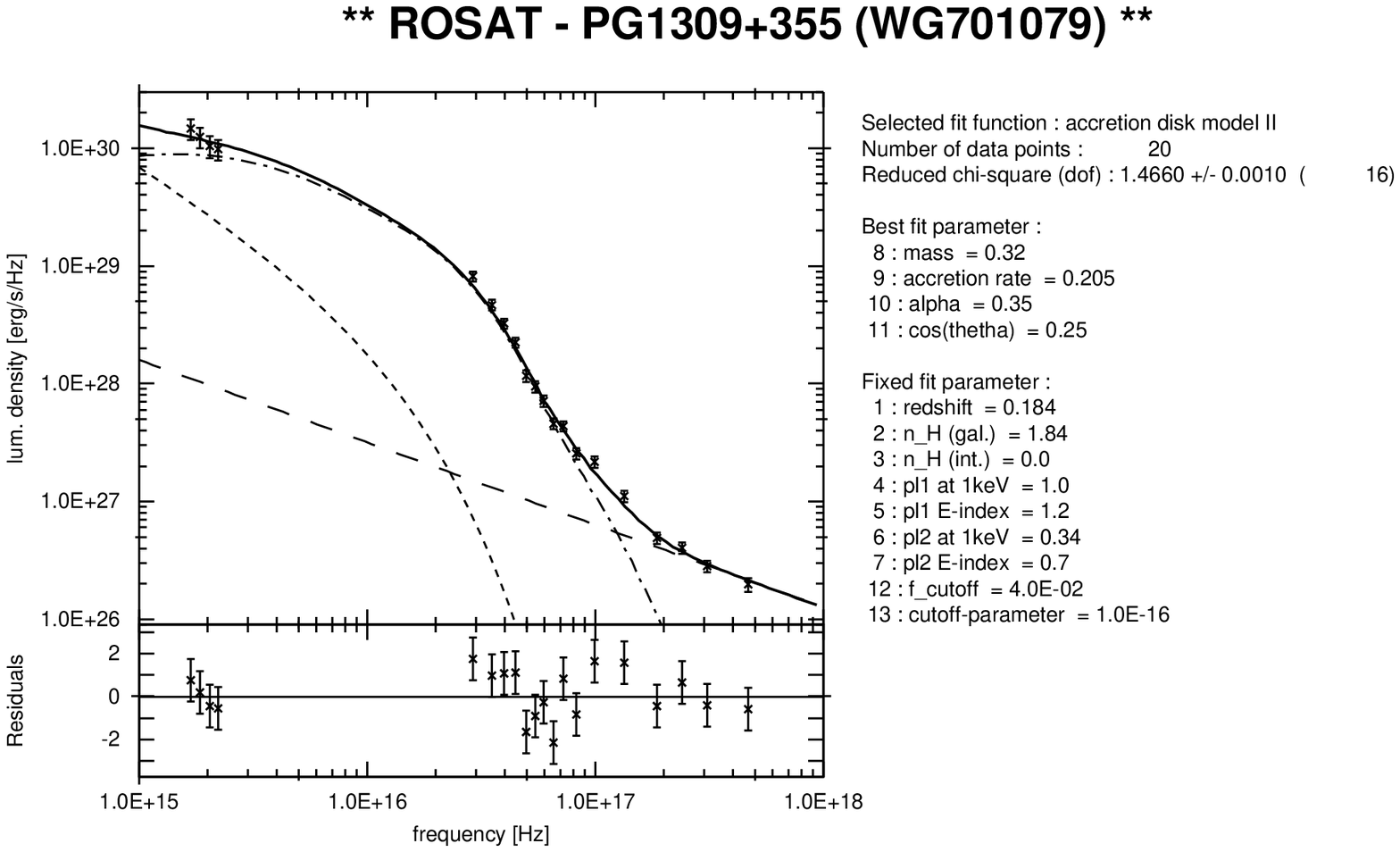,width=8.0cm,clip=}}

\par\centerline{\psfig{figure=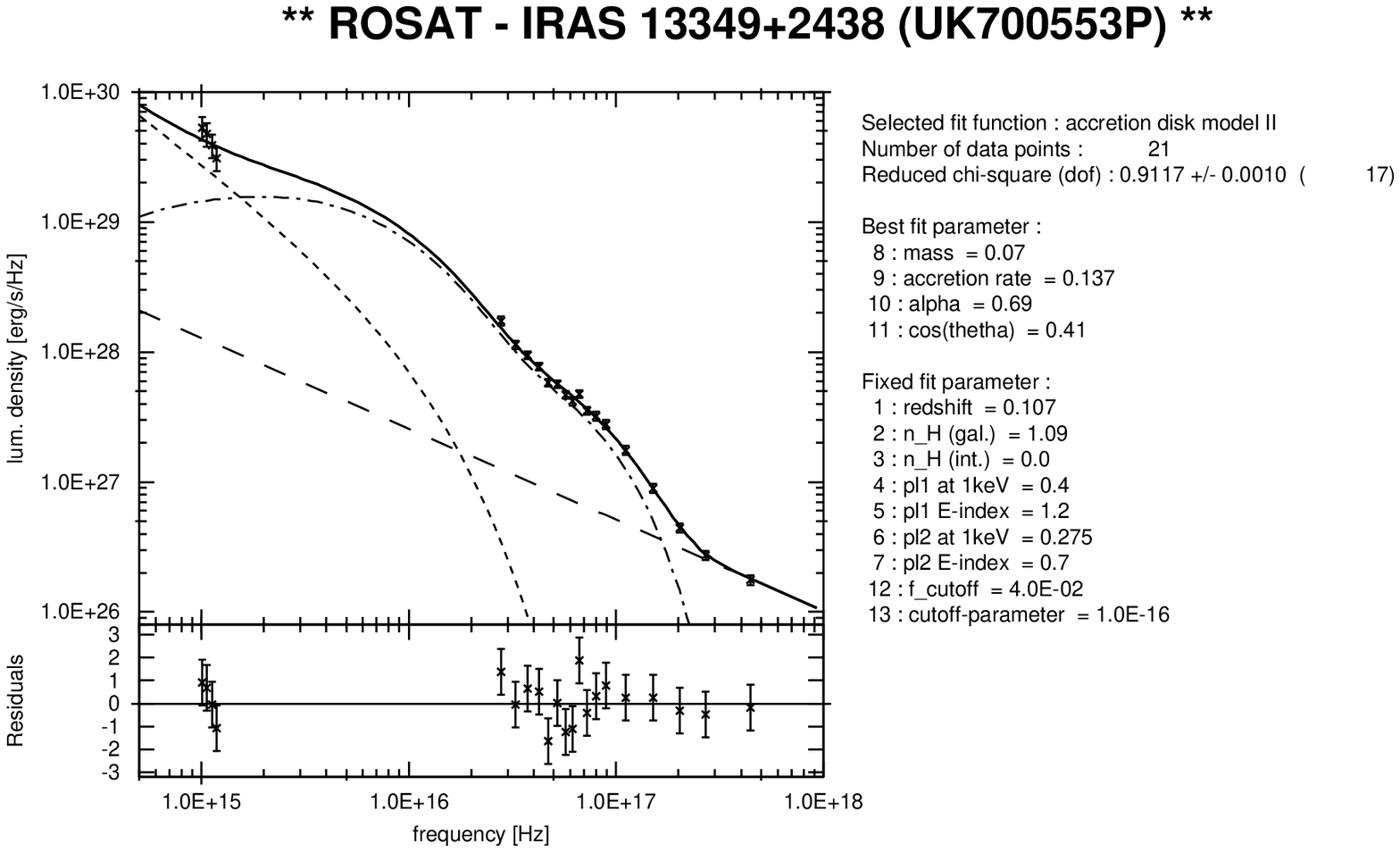,width=8.0cm,clip=}}
\caption[]{UV to X-ray spectra of sample members \#15, \#17, and \#18,
showing UV and ROSAT data points (crosses represent 1 $\sigma$ error bars) 
and model prediction (solid line). The accretion disk component (dot-dashed) 
and hard X-ray (long dashes) and IR (short dashes) power laws are 
displayed separately. The spectrum is plotted in the source frame. Residuals 
are given in units of $1 \sigma$.\label{spec}}
\end{figure}

\begin{figure}
\par\centerline{\psfig{figure=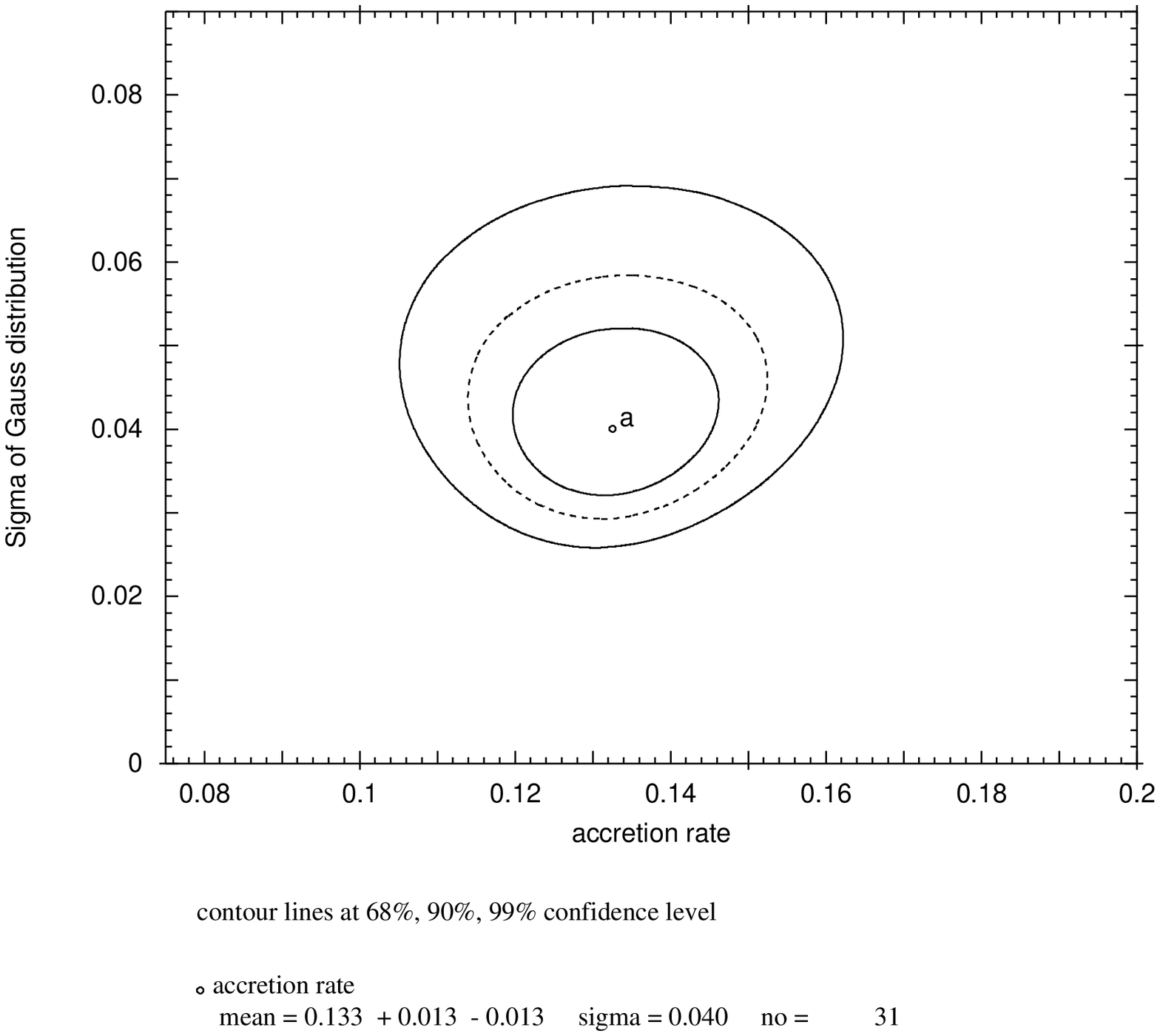,width=8.8cm,clip=}}
\caption[]{Distribution of best-fit accretion rates $\dot{M}$ in units of 
the Eddington accretion rate $\dot{M}_{\rm Eddington}$. Confidence contours 
(68 \%, 90 \%, 99 \%) of the mean and width (Gaussian $\sigma$) of the 
distribution are plotted.\label{contacc}}

\bigskip

\bigskip

\par\centerline{\psfig{figure=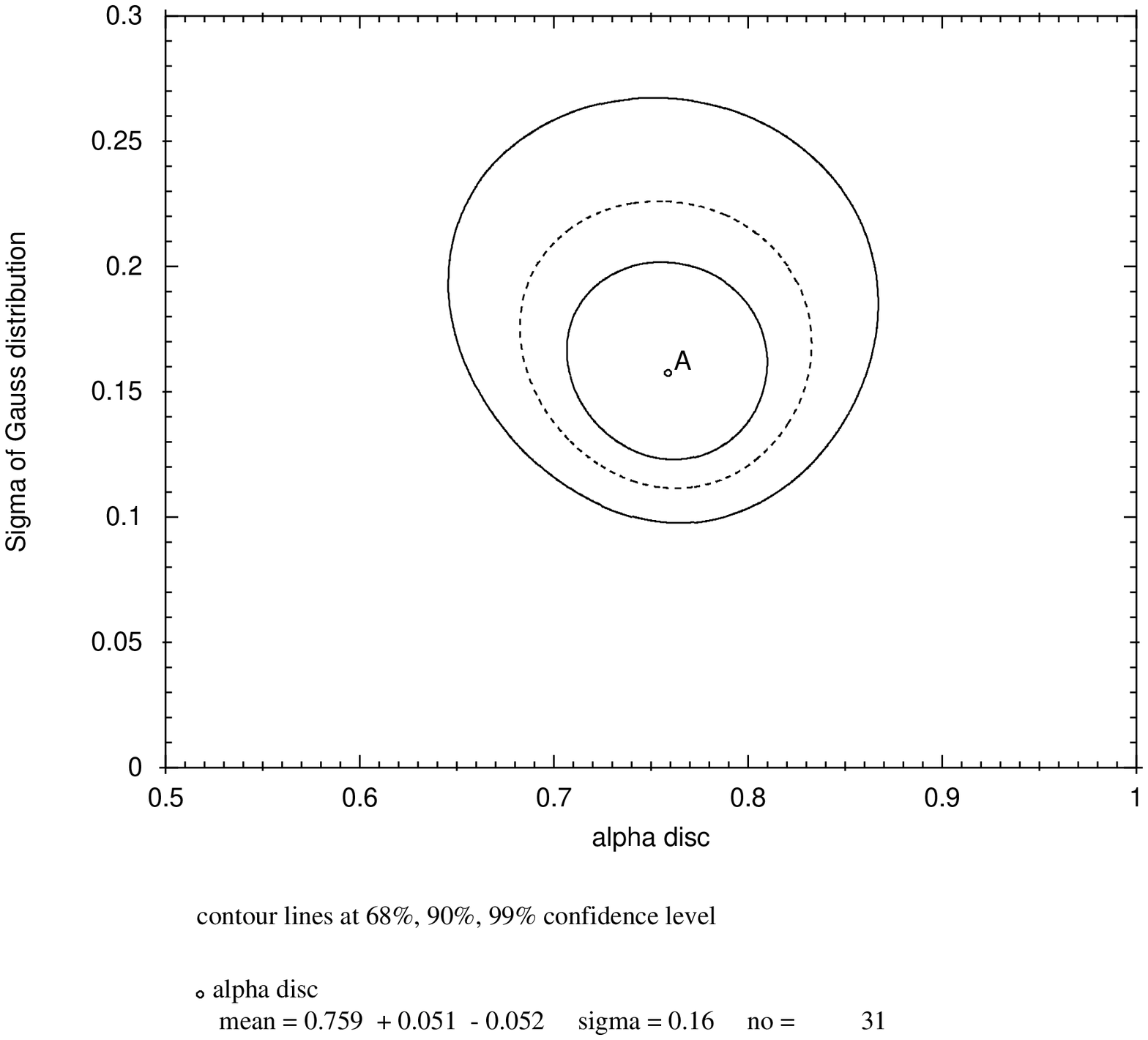,width=8.8cm,clip=}}
\caption[]{Distribution of best-fit viscosity parameters $\alpha$.
Confidence contours (68 \%, 90 \%, 99 \%) of the mean and width (Gaussian 
$\sigma$) of the distribution are plotted.\label{contalph}}
\end{figure}

As already mentioned in the introduction, in order to separate the
accretion disk emission from the underlying hard power law spectrum,
where available we took the spectral index for the hard power law from the 
literature ({\em Einstein Observatory}, EXOSAT, GINGA; Malaguti et al., 1994). 
Otherwise, a canonical value of $0.7$  was adopted. X-ray variability between 
the observation in the soft and hard X-ray bands was allowed for by treating 
the normalization of the hard power law spectrum as a free parameter. 
The spectral indices and references used are listed in Table \ref{fitpar}.

We have not included infrared and optical data in the spectral analysis, 
because such measurements were not available for all of the sample members
and many different emission components (non-thermal synchrotron emission, 
thermal dust emission, BLR and NLR line emission, 
emission from the host galaxy) would needlessly complicate the analysis.
Emission components at low frequencies which in addition to the
accretion disk also contribute to the emission in the UV range were 
incorporated into the model by assuming a power law component 
with spectral index $\alpha=1.2$ and an exponential cutoff in the extreme
UV range ($\nu =10^{16}$ Hz) which mainly represents the non-thermal 
synchrotron component. 

Interstellar low energy X-ray absorption is included in our
spectral fits by using the galactic hydrogen column densities by
Stark et al.\ \cite{stark} and Elvis et al.\ \cite{elvis} and by applying
the cross sections by Morrison \& Mc Cammon (\cite{mormc}).
Note that the de-reddening of the UV data, based on the same  
hydrogen column densities, was already applied to the data prior to the
spectral fitting (see section 2).
We did not consider intrinsic absorption in the accretion disk model fitting.
This is justified as no evidence for intrinsic absorption was found in 
our spectral power law fits and, generally, high luminosity objects 
(i.e. quasars as opposed to lower luminosity AGN) at moderate redshifts 
are not expected to show large amounts of intrinsic absorption.  

The total fit function thus includes three components with a total of six free
parameters:

1. Accretion disk model with four free parameters ($M$,$\dot{M}$,$\alpha$,
$\Theta _{0}$)

2. Underlying hard power law with a slope taken from the
literature (free normalization)

3. Soft power law with $\alpha =1.2$ and a cutoff in the EUV 
(free normalization).

The accretion rate $\dot{M}$ is measured in units of the 
Eddington accretion rate $\dot{M}_{\rm Edd}=L_{\rm Edd}/(\eta c^{2})$,
where $L_{\rm Edd}=4\pi c G M/\kappa _{T}$ is the Eddington luminosity,
$\eta$ is the efficiency of accretion, and $\kappa _{T}$ is the
Thomson opacity.
One additional parameter of the accretion disk model, the specific angular 
momentum $a/M$ of the black hole was fixed at $a/M=0$, i.e., we have 
considered only the non-rotating case with $\eta =0.057$.

To compare the accretion disk model with observations, we calculated model 
spectra for a fixed grid in 4 dimensional parameter space 
consisting of 980 data points (see Table \ref{para}).
At intermediate points within the grid, model spectra were computed by linear 
interpolation. 

\begin{table}[htb]
\caption[]{Grid of accretion disk model parameters}
\label{para}
\begin{tabular}{lcl}
\hline
{\bf parameter} & {\bf range}
                & \multicolumn{1}{c}{\bf parameter values}     \\ \hline \hline
$M [M_{\odot}]$     & [$10^7, 10^{10}$] 
                    & $10^{7}$, $10^{8}$, $10^{9}$, $10^{10}$  \\
$\dot{M} [\dot{M}_{\rm Edd}]$ & [0.1, 0.9]
                    & 0.1, 0.2, 0.3, 0.4, 0.5, 0.7, 0.9 \\ 
$\alpha $           & [0.1, 0.9] 
                    & 0.1, 0.2, 0.3, 0.4, 0.5, 0.7, 0.9 \\ 
$\cos \Theta _{0}$  & [0.0, 1.0]
                    & 0.0, 0.25, 0.5, 0.75, 1.0                \\ \hline
\end{tabular}
\end{table}

Simultaneous fits of the ROSAT and IUE data were performed by folding
the model X-ray fluxes with the response of the ROSAT PSPC detector
to determine the ROSAT count rates predicted by our model in each spectral 
bin. The UV fluxes predicted by the model were compared directly with
the de-reddened IUE continuum fluxes (see section 3).
Model fluxes were calculated from the source luminosities emitted
by the accretion disk by assuming a Hubble constant 
$H_0 = 50 \; {\rm km} \; {\rm s}^{-1} \; {\rm Mpc}^{-1}$
and $q_0 = 0$. Note that in particular the best-fit central masses $M$ are 
dependent on these assumptions, i.e., $M$ is approximately proportional to 
$H_0^{-2}$. Statistical errors of the best-fit parameters were determined by
calculating a $\chi^2$ grid in parameter space. The region defined by
$\chi^2_{\rm min}[11~ + 4.72$, corresponding to the 68 \% value of the cumulative
$\chi^2$-distribution for 4 degrees of freedom, was then used to construct
the upper and lower 1 $\sigma$ errors of the fit parameters. 

\begin{table*}
\caption[ ]{\label{fitpar}Table of the resulting fit parameters. \newline
$\alpha_{hard}$: Spectral index of the underlying hard power law. 
$^{\dagger}$ mean value of Exosat and Ginga observations; $^{\ddagger}$ 
additional HEAO data. All data are taken from Malaguti et al. \cite{malaguti}.
$\alpha_{hard}$ was set to 0.7 where no measurements were available.
All parameters are given with their upper ($+$) and lower ($-$) 1 
$\sigma$-errors. If there is no value in the list (-) the error exceeds 
the parameter boundary. Masses are in units of $10^9 M_\odot$ and accretion
rates are in units of the Eddington accretion rate.}
\begin{flushleft}
\begin{tabular}{l|l|ccc|ccc|ccc|ccc|c}
\noalign{\smallskip}
\hline
\noalign{\smallskip}
name & $\alpha_{hard}$ & $M$ & $- \sigma$ & $+ \sigma$ & $\dot{M}$ 
& $- \sigma$ & $+ \sigma$ & $\alpha_{visc.}$ 
&  $- \sigma$ &  $+ \sigma$
& cos $\vartheta$ & $- \sigma$ & $+ \sigma$ & $ \chi ^2$\\
\noalign{\smallskip}
\hline \hline
\noalign{\smallskip}
0026$+$129 & 0.91$^{\dagger}$  & 1.001 & 0.768 & 2.556 & 0.1034 & 0.0015 
& 0.0038 & 0.904 & 0.120 & -     & 0.20 & -    & -    & 0.34 \\
0052$+$251 & 0.95$^{\dagger}$  & 1.005 & 0.692 & 3.795 & 0.1092 & 0.0016 
& 0.0067 & 0.793 & 0.036 & 0.098 & 0.30 & -    & -    & 0.73 \\
0119$-$286 & 0.70              & 0.470 & 0.297 & 0.751 & 0.1287 & 0.0081 
& 0.0120 & 0.801 & 0.017 & 0.033 & 0.29 & 0.22 & 0.56 & 2.36 \\
0157$+$001 & 0.70              & 0.291 & 0.058 & 3.457 & 0.1064 & -      
& 0.0012 & 0.886 & 0.107 & 0.046 & 0.94 & -    & -    & 0.67 \\
0205$+$024 & 0.70              & 0.343 & 0.131 & 1.178 & 0.1331 & 0.0111 
& 0.1064 & 0.807 & 0.018 & 0.033 & 0.54 & 0.47 & 0.32 & 1.47 \\
0804$+$761 & 1.04$^{\dagger}$  & 0.240 & 0.056 & 2.798 & 0.1089 & -      
& 0.0027 & 0.870 & 0.082 & 0.052 & 0.91 & -    & -    & 0.66 \\
0914$-$621 & 0.70              & 0.378 & 0.184 & 3.196 & 0.1471 & 0.0369 
& 0.1422 & 0.593 & 0.090 & 0.145 & 0.64 & -    & -    & 1.22 \\
1029$-$140 & 0.70              & 1.250 & 0.498 & 0.488 & 0.1147 & 0.0019 
& 0.0018 & 0.812 & 0.010 & 0.015 & 0.27 & 0.11 & 0.25 & 3.53 \\
1049$-$005 & 0.70              & 1.906 & 1.495 & 7.582 & 0.1112 & 0.0089 
& 0.1285 & 0.601 & -     & 0.280 & 0.27 & -    & -    & 1.07 \\
1100$-$264 & 0.70              & 9.201 & 2.762 & -     & 0.2028 & -      
& -      & 0.755 & -     & -     & 1.00 & 0.39 & -    & 1.20 \\
1116$+$215 & 1.00$^{\ddagger}$ & 1.854 & 1.155 & 0.617 & 0.1073 & 0.0011 
& 0.0013 & 0.838 & 0.016 & 0.018 & 0.27 & 0.08 & 0.36 & 3.44 \\
1202$+$281 & 0.70              & 0.699 & -     & 0.154 & 0.1066 & 0.0017 
& 0.0034 & 0.920 & 0.057 & 0.049 & 0.12 & -    & -    & 1.07 \\
1216$+$069 & 0.70              & 0.612 & 0.143 & -     & 0.1123 & 0.0065 
& 0.0325 & 0.978 & 0.144 & -     & 0.96 & -    & -    & 0.88 \\
1247$+$267 & 0.70              & 8.997 & 4.448 & -     & 0.2450 & 0.0945 
& -      & 0.471 & 0.253 & 0.485 & 0.74 & 0.35 & -    & 1.05 \\
1257$+$346 & 0.70              & 4.900 & 2.128 & -     & 0.1185 & 0.0087 
& 0.0218 & 0.947 & -     & -     & 0.76 & 0.47 & -    & 0.75 \\
1307$+$085 & 1.08$^{\ddagger}$ & 0.252 & -     & -     & 0.1103 & 0.0051 
& 0.0001 & 0.784 & 0.119 & 0.046 & 1.00 & -    & -    & 1.66 \\
1309$+$355 & 0.70              & 0.321 & 0.215 & 1.509 & 0.2052 & 0.0044 
& 0.0064 & 0.350 & 0.045 & 0.042 & 0.25 & 0.25 & -    & 1.47 \\
1334$+$246 & 0.70              & 0.073 & -     & 0.591 & 0.1374 & 0.0117 
& 0.0629 & 0.689 & 0.050 & 0.090 & 0.41 & -    & -    & 0.91 \\
1352$+$183 & 0.70              & 0.299 & 0.157 & 1.756 & 0.1100 & -      
& 0.0043 & 0.786 & 0.044 & 0.091 & 0.44 & -    & -    & 0.75 \\
1407$+$265 & 0.70              & 9.261 & 6.149 & -     & 0.2861 & 0.0166 
& 0.0190 & 0.546 & 0.039 & 0.027 & 0.11 & -    & 0.42 & 1.04 \\
1415$+$451 & 0.70              & 0.088 & 0.051 & 0.187 & 0.1121 & 0.0024 
& 0.0106 & 0.650 & 0.049 & 0.061 & 0.50 & 0.43 & -    & 2.07 \\
1416$-$129 & 0.51$^{\dagger}$  & 0.210 & -     & 0.545 & 0.1138 & 0.0019 
& 0.0048 & 0.987 & 0.032 & -     & 0.50 & 0.42 & 0.25 & 0.76 \\
1444$+$407 & 0.70              & 0.497 & 0.203 & -     & 0.1135 & 0.0066 
& -      & 0.903 & 0.062 & 0.073 & 0.66 & -    & -    & 1.20 \\
1521$+$101 & 0.70              & 4.257 & 1.320 & -     & 0.2141 & 0.1043 
& 0.1890 & 0.483 & 0.162 & -     & 1.00 & 0.72 & -    & 1.10 \\
1543$+$489 & 0.70              & 0.930 & 0.360 & 8.968 & 0.1081 & 0.0043 
& 0.0046 & 0.925 & 0.060 & -     & 0.72 & -    & -    & 0.92 \\
1613$+$658 & 0.70              & 0.195 & 0.028 & 2.262 & 0.1210 & 0.0111 
& -      & 0.863 & 0.073 & 0.033 & 0.91 & -    & -    & 0.50 \\
1617$+$175 & 0.70              & 0.493 & 0.346 & 2.607 & 0.1410 & -      
& 0.0734 & 0.304 & -     & 0.141 & 0.28 & -    & -    & 1.57 \\
1630$+$377 & 0.70              & 2.639 & 0.595 & -     & 0.2136 & 0.0067 
& 0.0096 & 0.585 & 0.178 & 0.061 & 1.00 & 0.83 & -    & 1.19 \\
1700$+$642 & 0.70              & 5.869 & 2.411 & -     & 0.2003 & 0.0991 
& -      & 0.794 & -     & -     & 0.84 & 0.54 & -    & 1.35 \\
1821$+$643 & 0.89$^{\dagger}$  & 2.092 & 0.861 & -     & 0.1990 & 0.0681 
& 0.0029 & 0.562 & 0.035 & -     & 0.60 & 0.56 & -    & 3.97 \\
2251$-$178 & 0.53$^{\dagger}$  & 0.251 & 0.117 & 0.255 & 0.1227 & 0.0041 
& 0.0044 & 0.816 & 0.011 & 0.010 & 0.35 & 0.21 & 0.43 & 2.76 \\
\noalign{\smallskip}
\hline
\noalign{\smallskip}
\end{tabular}
\end{flushleft}
\end{table*}

\begin{figure}
\par\centerline{\psfig{figure=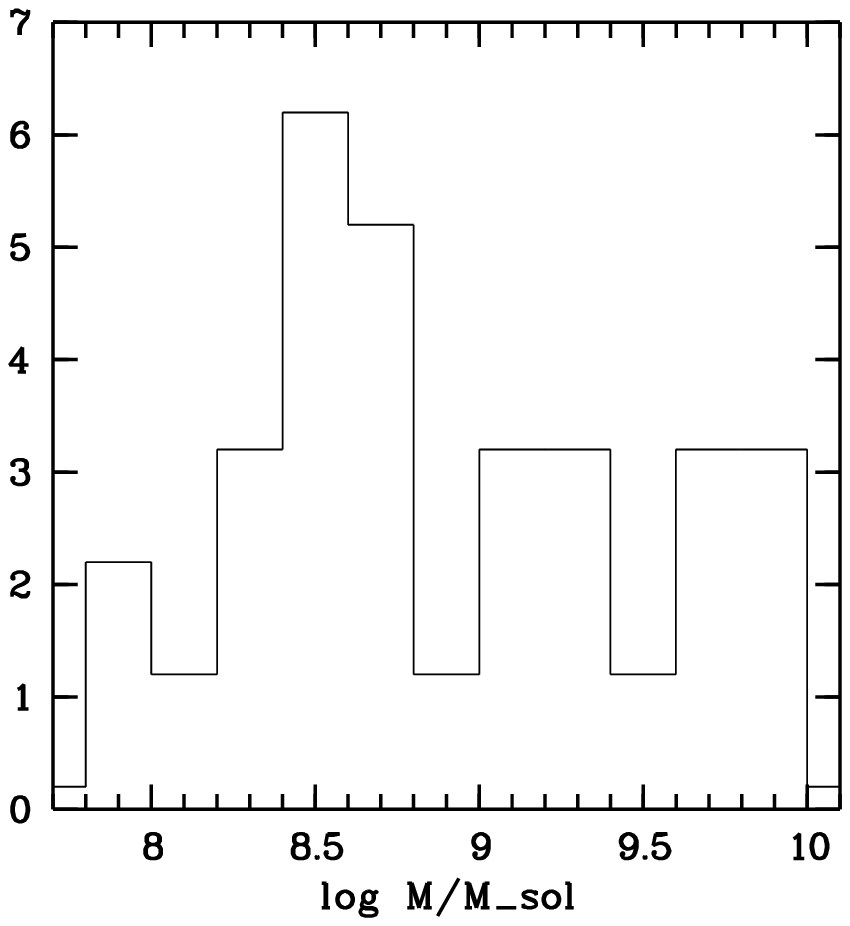,width=8.8cm,clip=}}
\caption[]{Histogram of best-fit central masses, $M$.\label{mass}}
\par\centerline{\psfig{figure=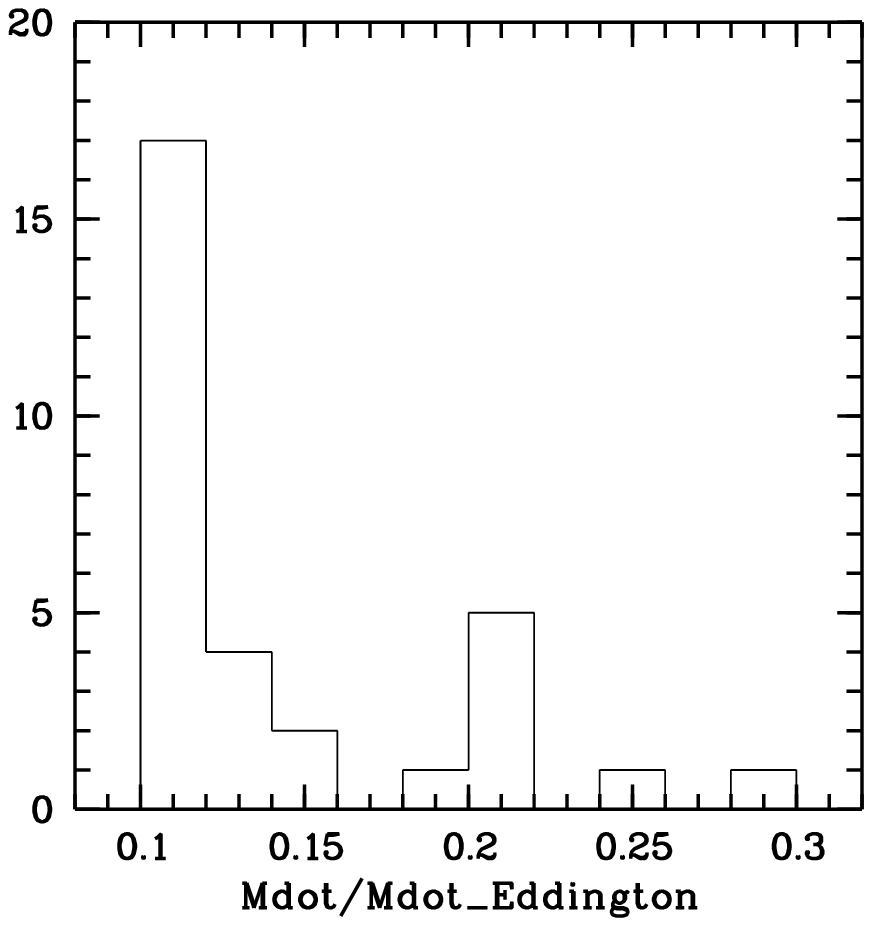,width=8.8cm,clip=}}
\caption[]{Histogram of best-fit mass accretion rates, $\dot{M}$, in
units of the Eddington accretion rate, $\dot{M}_{\rm Eddington}$.
\label{mdothist}}
\end{figure}

\begin{figure*}
\par\centerline{\psfig{figure=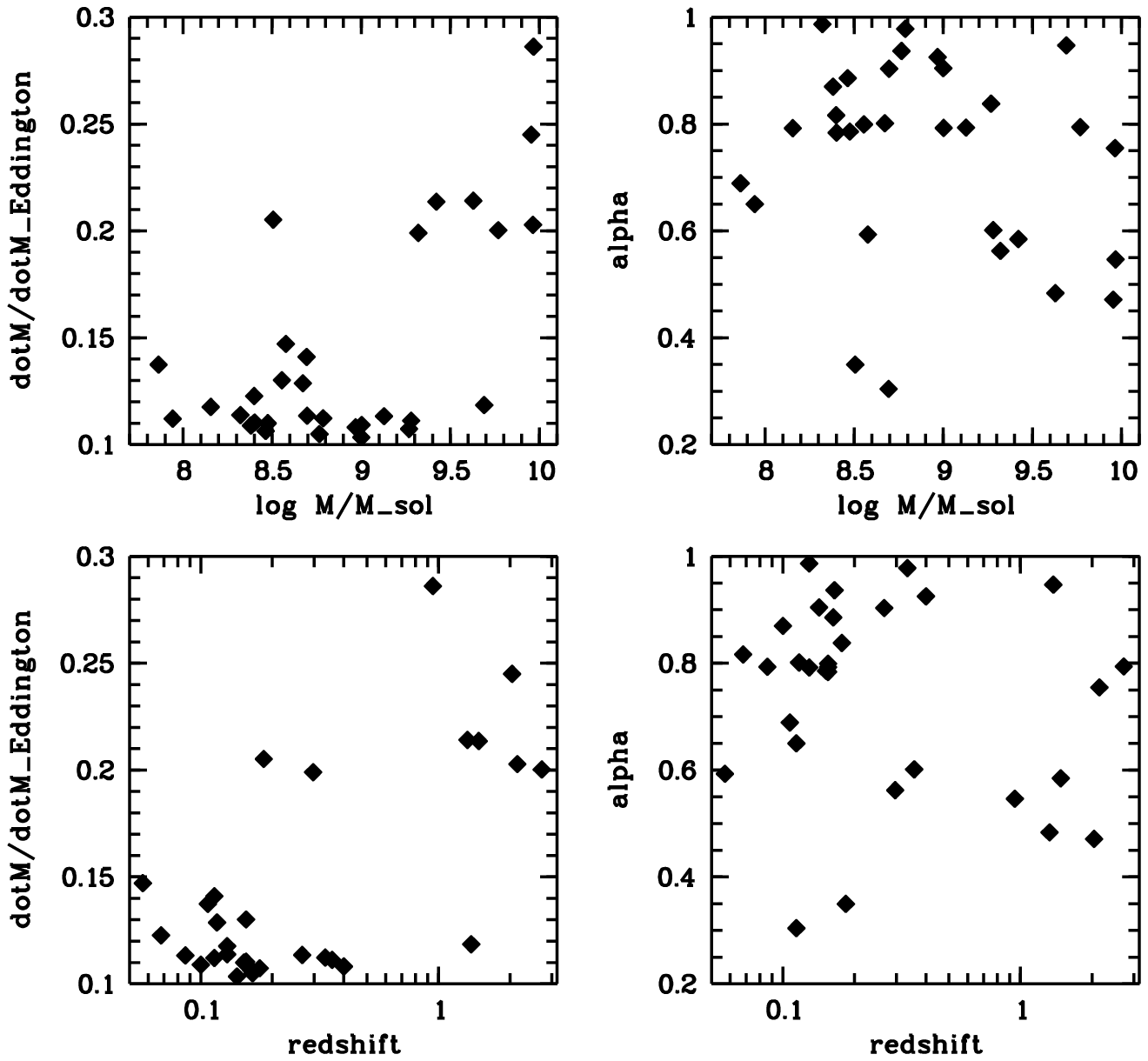,width=15cm,clip=}}
\caption[]{Best-fit accretion rates in units of the Eddington
accretion rate and viscosity parameters plotted over best-fit central masses
and redshifts.\label{params}}
\end{figure*}

\section{Results of the model fitting}

As examples, we present in Fig. \ref{spec} the best fit spectra of three 
sample members, \#15, \#17, and \#18 (Tab. \ref{sample}), showing both the 
IUE and ROSAT data points as well as the prediction from our model 
calculation. Due to the complex nature of the spectrum in the UV range, 
in some objects the slope of the UV continuum is not well
matched by our model fits (see, for example, object \# 18 in Fig. \ref{spec}).
This has no large effect on the best-fit $\chi^2$ values, however, which 
are dominated by the X-ray data points, while the UV data points 
mainly help to constrain the total flux from the accretion disk.

The resulting best fit parameters of all sample members, as well as
their $1 \sigma$ errors and the corresponding minimum $\chi ^{2}$ values are 
given in Table \ref{fitpar}. Cases where the upper or lower errors lie 
beyond the limit of our calculated grid, are denoted by a minus sign. In most 
cases acceptable fits are achieved. 
The distribution of model parameters was studied
using a maximum likelihood technique (see Avni 1976) which, based
on the assumption that both the model parameters and their statistical
errors follow the normal distribution, gives the first (mean) and second 
moment (i.e., the $\sigma$ of the normal distribution) as well as their 
statistical errors. Figure \ref{contacc} and \ref{contalph} show the 
68 \%, 90 \%, and 99 \% confidence contours of the distribution of $\dot{M}$ 
and $\alpha$, respectively. 
The case that all sample members have the same $\dot{M}$ and/or $\alpha$ 
best-fit parameter values is excluded at a high statistical significance level
(the confidence contours do not intersect the $\sigma=0$ line).
We find a mean accretion rate of $<\dot{M}>\; = 0.13 \; 
\dot{M}_{\rm Eddington}$ within a relatively narrow parameter range ($\sigma 
\approx 0.05 \; \dot{M}_{\rm Eddington}$). The best-fit accretion rates 
$\dot{M}$
are below $0.3 \; \times$ the Eddington accretion rate in all sample members
(see Fig. \ref{mdothist}), thus fulfilling the requirement for the thin disk 
approximation (Laor \& Netzer, 1989). The viscosity parameters are relatively 
high ($<\alpha >\; = 0.76$) and are spread over a wider range 
($0.5 < \alpha < 1.0$ for most objects), possibly suggesting some
diversity of the underlying physical viscosity mechanism in our sample.
Note that, according to its definition, $\alpha$ should not greatly 
exceed unity. 

The best-fit central masses which roughly span two orders of magnitude
($10^8 M_\odot - 10^{10} M_\odot$; see Fig. \ref{mass}) are in broad 
agreement with AGN black hole masses derived from variability and from general 
luminosity arguments.
As the accretion rates are found to be confined within a relatively
narrow range ($\dot{M} = 0.1 - 0.3 \times {M}_{\rm Eddington}$), 
this implies that, in absolute terms, the mass accretion rates also span
about two orders of magnitude while maintaining a rough proportionality
(within a factor $\sim 3$) with the central masses over the whole dynamic 
range. We have tested for any dependencies of $\dot{M}$ and $\alpha$ on $M$ 
and find that low central masses also seem to be associated with accretion at 
a lower fraction of the Eddington accretion rate, 
$\dot{M}/\dot{M}_{\rm Eddington}$: When $M$ is increased by two orders of 
magnitude (from $10^8 M_\odot$ to $10^{10} M_\odot$) a moderate increase of 
$\dot{M}/\dot{M}_{\rm Eddington}$ by roughly a factor of three 
(from 0.1 to 0.3) 
is observed. No such dependence of the viscosity parameter 
$\alpha$ on central mass is observed. Scatter plots of $\dot{M}$ 
and $\alpha$ plotted over $M$ and redshift are shown in Fig. \ref{params}. 
High central masses also imply higher luminosities and, on average, 
larger distances. Any dependence on central mass thus is also expected to 
result in a similar dependence on redshift, as is observed (Fig. 
\ref{params}, lower panel). 
Note that the observed dependencies of $\dot{M}/\dot{M}_{\rm Eddington}$
on central mass and redshift can not be attributed to selection effects, 
alone: Objects with, e.g., central masses of $8 \times 10^8 M_\odot$ at a 
redshift of $z=0.1$ would be well above the respective X-ray and UV 
sensitivity limits if their mass accretion rates were higher than the 
observed values $< 0.15 \; \dot{M}_{\rm Eddington}$. 

The narrow range of observed accretion rates in terms of the Eddington
accretion rate also implies that the large luminosity range covered by
AGN must predominantly be due to a similarly large variation in central mass. 
Note that for the object class  studied here, i.e. radio-quiet quasars, 
the emission is considered to be dominated by 
an unobscured accretion disk and absorbing material on the line of sight
is thus not thought to contribute to the large observed luminosity range.
Taken together with the known evolution of the quasar luminosity function 
(e.g., Boyle et al., 1987 and 1993), i.e. the fact that 
quasars at high redshifts are considerably more luminous than `local' 
quasars (by up to a factor of 40 at redshift z = 2, depending on which 
relative contribution of luminosity and/or density evolution is favoured) 
it follows that quasars at earlier epoches were more massive than present 
day quasars by similar factors, giving further support to the concept that 
many local galaxies (including our own; Genzel \& Eckart, \cite{genzel}) 
contain dormant, super-massive black holes in their centers. See the more 
de 4 tailed discussion of this finding in Brunner et al. (1997). 
 
We presently do not know which physical processes are responsible
for the fact that high accretion rates (0.3 -- 1.0 $\dot{M}_{\rm Eddington}$ 
for the total sample; 0.15 -- 1.0 $\dot{M}_{\rm Eddington}$ for the low 
mass/low 
redshift subsample) are not observed. However, since the definition of the 
Eddington accretion rate is based on the assumption that both radiation and 
accretion flow are isotropic, suitable unisotropies of both the accretion flow
and the resulting radiation may lead to a reduction of the permitted maximum 
accretion rates. Dynamical processes in the disk not considered in our
present modeling may also result in a limit to the possible accretion rates.
We believe that this highly interesting point warrants further
theoretical attention. Note that the observed lower cutoff of the 
distribution of accretion rates ($\dot{M} \geq 0.1 \; \dot{M}_{\rm Eddington}$)
may be due to selection effects: At very low accretion rates no 
appreciable emission is expected in the X-ray range such that most objects 
will not be detected in the ROSAT band.

\begin{figure}
\par\centerline{\psfig{figure=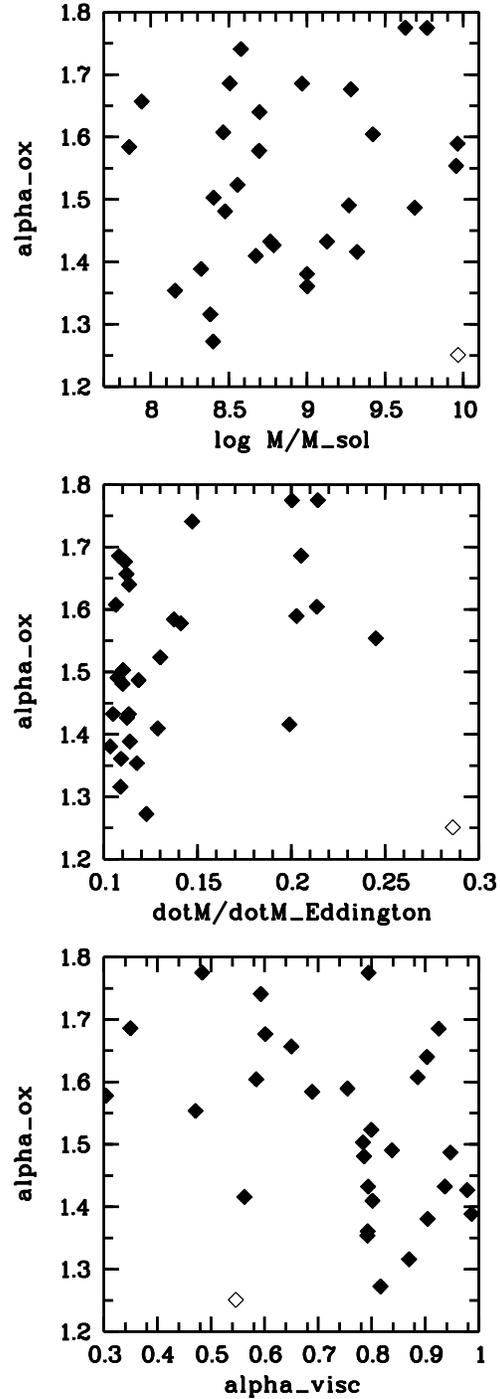,width=8.8cm,clip=}}
\caption[]{$\alpha_{\rm ox}$ plotted over accretion disk parameters, 
central mass $M$, accretion rate $\dot{M}/\dot{M}_{\rm Eddington}$, and 
viscosity parameter $\alpha_{\rm visc.}$. One anomalous object with the 
lowest $\alpha_{\rm ox}$ value in the sample is plotted as an open dimond. 
\label{aoxdisk}}
\end{figure}

We find marginal correlations of the accretion disk parameters with 
$\alpha_{\rm ox}$ (see Fig. \ref{aoxdisk}). When one anomalous object, \# 20
(Table \ref{sample}), with the lowest $\alpha_{\rm ox}$, and at the same time 
the highest $M (9.26 \cdot 10^9 M_\odot)$ and $\dot{M} \; (0.286 \;
\dot{M}_{\rm Eddington}$) values in the sample is removed, the probabilities
for randomness of the observed correlations of $\alpha_{\rm ox}$ on $M$, 
$\dot{M}/\dot{M}_{\rm Eddington}$, and $\alpha$ are 0.16, 0.04, and 0.02,
respectively (Spearman rank correlation). Note that both $M$ and 
$\dot{M}/\dot{M}_{\rm Eddington}$ contribute to the total luminosity of 
the disk
($L \sim M \times \dot{M}/\dot{M}_{\rm Eddington}$). Since in the optical 
spectral
range the emission is dominated by the accretion disk while at X-ray energies
a large fraction of the emission is supplied by the hard power law component,
an increase of either $M$ or $\dot{M}/\dot{M}_{\rm Eddington}$ predominantly 
affects the optical emission and thus results in an increase of the 
broad-band spectral index $\alpha_{\rm ox}$, in agreement with the observed 
correlations. By increasing the viscosity parameter $\alpha$ a larger part 
of the disk emission is radiated in the X-ray range, thus resulting in a 
hardening of the broad-band spectral index $\alpha_{\rm ox}$, again in 
agreement with observations.

A statistical comparison of the sample properties of AGN from the 
ROSAT All Sky Survey using a simpler precursor version of the present 
accretion disk code has been performed by Friedrich et al. (\cite{friedrich})
which is in broad agreement with the present study. However, using the 
improved model, considerably smaller mass accretion rates are sufficient to 
produce the observed X-ray emission. 
This is mainly because, contrary to the simpler version, our improved 
model also takes into account the temperature gradient in the 
vertical direction of the disk. This means that the local spectra differ from
the blackbody even in the optically thick case, leading to harder spectra 
for the same parameter values.

\section{Concluding remarks}

We are aware that the accretion disk model fitting performed leaves many 
open ends. I.e., contributions from other ingredients of the AGN 
phenomenon such as non-stationary processes in the disk, illumination of 
the disk, effects on the spectrum due to 
reflection and due to an absorbing torus have been neglected. 
However, the fact that in most cases acceptable fits of our accretion disk 
model were achieved has strengthend our belief that, for the objects 
selected, to first order this approach is justified. We thus have
demonstrated that the `naked' (excluding the above modifications) 
accretion disk model is a viable, and in fact we think the most promising 
candidate, for understanding the soft X-ray excess and big blue 
bump emission in at least some classes of AGN (radio-quiet quasars and Seyfert
I galaxies). Further modeling including some of the above ingredients will, 
however, be highly desirable, particularly when considering the availability 
before the end of the century of X-ray instruments with much improved  
sensitivity, spectral resolution, and spectral coverage.  

\begin{acknowledgements}
This work was supported by DARA under grants 50 OR 9009, 50 OR 9603, and
50 QR 8802. It has made use of the IUE/ULDA (ESA), NED (NASA/IPAC 
Extragalactic Database), and HEASARC (High Energy Astrophysics Science 
Archive Research Center) online databases.
\end{acknowledgements}

{}


\begin{thebibliography}{}
\bibitem[1987]{bechtold}
Bechtold, J., Czerny, B., Elvis, M., Fabiano, G., Green, R.\ F., 
1987, ApJ, 314, 699
\bibitem[1987]{bohlin}
Bohlin, R.\  C., Savage, B.\ D., Drake, J.\ F., 1978, ApJ, 224, 132 
\bibitem[1987]{boyle87}
Boyle, B.\ J., Fong, R., Shanks, T., and Peterson, B.\ A., 1987,
MNRAS, 227, 717
\bibitem[1993]{boyle93}
Boyle, B.\ J., Griffiths, R.\ E., Shanks, T., Stewart, G.\ C., 
Georgantopoulos, I., 1993, MNRAS, 260, 49
\bibitem[1997]{brunn}
Brunner, H., Lamer, G., D\"orrer, T., Friedrich, P., Staubert, R., 
1997, A\&A, in preparation
\bibitem[1975]{cunningham}
Cunningham, C.\ T., 1975, ApJ, 202, 788
\bibitem[1987]{czerny}
Czerny, C.\ T., Elvis, M., 1987, ApJ, 321, 305
\bibitem[1996]{doerrer}
D\"orrer, T., Riffert, H., Staubert, R., Ruder, H., 1996, 
A\&A, 311, 69 
\bibitem[1986]{edelson}
Edelson, R.\ A., Malkan, M.\ A., 1986, ApJ, 308, 59.
\bibitem[1989]{elvis}
Elvis, M., Lockman, F.\ J., Wilkes, B.\ J., 1989, AJ, 97, 777
\bibitem[1997]{friedrich}
Friedrich, P., D\"orrer, T., Brunner, H., Staubert, R., 1997, A\&A, 
submitted
\bibitem[1996]{genzel}
Genzel, R., Eckart, A., 1996, Nature, 383, 415
\bibitem[1993]{hewitt}
Hewitt, A., Burbidge, G., 1993, ApJS, 87, 451
\bibitem[1957]{kompaneets}
Kompaneets, A.\ S., 1957, Soviet Phys.\ JETP, 4, 730
\bibitem[1989]{laor89}
Laor, A., Netzer, H., 1989, MNRAS, 238, 897
\bibitem[1990]{laor90}
Laor, A., Netzer, H., Piran, T., 1990, MNRAS, 242, 560
\bibitem[1994]{malaguti}
Malaguti, G., Bassani, L., Caroli, E., 1994, ApJS, 94, 517
\bibitem[1982]{malkan}
Malkan, M.\ A., Sargent, W.\ L.\ W., 1982, ApJ, 254, 22
\bibitem[1983]{malkan2}
Malkan, M.\ A., 1983, ApJ, 268, 582
\bibitem[1983]{marshall}
Marshall, H.\ L., 1983, ApJ, 269, 42
\bibitem[1983]{mormc}
Morrison, R.\ McCammon, D., 1983, ApJ, 270, 119
\bibitem[1973]{novikov}
Novikov, I.\ D., Thorne, K.\ S., 1973, in {\it Black Holes},
eds.\ de Witt, C., de Witt, B., Gordon \& Breach, New York
\bibitem[1996]{puch96}
Puchnarewicz, E.\ M., Mason, K.\ O., Romero-Colmenero, E., et al., 1996,
MNRAS, 281, 1243
\bibitem[1995]{riffert}
Riffert, H., Herold, H., 1995, ApJ, 450, 508
\bibitem[1992]{ross}
Ross, R.\ R., Fabian, A.\ C., Mineshige, S., 1992, MNRAS, 258, 189
\bibitem[1996]{schartel96}
Schartel, N., Green, P.\ J., Anderson, S.\ F., et al. 1996, MNRAS, 
283, 1015
\bibitem[1979]{seaton}
Seaton, M.\ J., 1979, MNRAS, 187, 73
\bibitem[1973]{shakura}
Shakura, N.\ I., Sunyaev, R.\ A., 1973, A\&A, 24, 337
\bibitem[1993]{shimura}
Shimura, T., Takahara, F., 1993, ApJ, 419, 78
\bibitem[1995]{shimura2}
Shimura, T., Takahara, F., 1995, ApJ, 440, 610
\bibitem[1995]{speith}
Speith, R., Riffert, H., Ruder, H., 1995, Comp.\ Phys.\ Comm., 88, 109
\bibitem[1992]{stark}
Stark, A.\ A., Gammie, C.\ F., Wilson, R.\ W., et al., 1992, 
ApJS, 79, 77
\bibitem[1992]{staubert}
Staubert, R., D\"orrer, T., M\"uller, C., Friedrich, P.,
Brunner, H., 1997, in {\it Lecture Notes in Physics}, eds., Spruit, H.\
and Meyer-Hoffmeister, E. 
\bibitem[1993]{veron}
Veron-Cetty, M.\ P., Veron, P., 1993, ESO, Sci. Rep., 13, 1 
\bibitem[1988]{wandel}
Wandel, A., Petrosian, V., 1988, ApJ, 329, L11
\bibitem[1981]{zamorani}
Zamorani, G., Henry, J.\ P., Maccacaro, T., et al., 1981, ApJ, 245, 357
\end{thebibliography}
\end{document}